# Optimization of a FLASH Proton Beam Using a Track-Repeating Algorithm

Qianxia Wang[1,2,*], Uwe Titt[2], Radhe Mohan[2], Fada Guan[2,3], Yao Zhao[2], Ming Yang[2], Pablo Yepes[1,2,*]

[1] Department of Physics and Astronomy, MS 315, Rice University, 6100 Main Street, Houston, TX 77005, United States of America,

[2] Department of Radiation Physics, Unit 1420, The University of Texas MD Anderson Cancer Center, 1515 Holcombe Blvd., Houston, TX 77030, United States of America

3 Department of Therapeutic Radiology, Yale School of Medicine, 35 Park Street, New Haven, CT, 06511, United States of America

[*] Authors to whom any correspondence should be addressed

E-mail: qw14@rice.edu and yepes@rice.edu

**Purpose:** To optimize characteristics of proton beams of two different energies (86.4 MeV and 159.5 MeV) of the synchrotron at the University of Texas MD Anderson Proton Therapy Center for FLASH delivery using the Fast Dose Calculator (FDC), a track-repeating Monte Carlo algorithm.

**Methods:** A phase space file in a plane at 202 mm downstream of the beam exit window is generated through tuning parameters to match FDC results with measured or MCNPX Monte Carlo-simulated integrated depth-dose distribution (IDD) and lateral dose profiles. To spread out the Bragg peak, widen the beam and reduce the penumbra, a ridge filter (RF), a high-Z material scatterer and a collimator with compensator are inserted in the beam path and their shapes and sizes have been optimized. The FDC calculations are validated by comparing Geant4 Monte Carlo simulations. In addition, a set of algorithms to automatically choose the optimum dimensions of the beam shaping elements is developed and tested using the same beams. At the last part, dose rates for optimized beams were estimated by scaling their dose distributions to that of their original beams.

**Results:** The optimized 86.4 MeV beam had an 8.5 mm wide spread-out Bragg peak (SOBP) (proximal 90% to distal 90% of the maximum dose), 14.5 mm, 12.0 mm and 11.0 lateral widths with dose above 50%, 80% and 90% respectively and a 2.5 mm penumbra from 80% to 20% in the lateral profile for the energy. The 159.5 MeV beam had a SOBP of 39.0 mm and the lateral widths with dose above 50%, 80% and 90% of 20.5 mm, 15.0 and 12.5 mm when the source to surface distance (SSD) was 550 mm. Wider lateral widths was obtained with increased SSD. The FDC calculations had passing rates higher than 96% using 3mm/3% as the gamma-index criterion comparing with Geant4 simulations for both energies. The set of automatic algorithms can choose the proper dimensions for the high-density scatterer, RF, collimator and compensator efficiently. And the ….

**Conclusion:** The synchrotron generated proton beams can be modified for FLASH applications with suitably selected and optimized high-Z scatterers, RFs, collimators and compensators. The track-repeating Monte Carlo algorithm FDC can accurately predict dose distributions when beam shaping elements of wide range of materials are included. The set of automatic algorithms work efficiently in optimizing the dimensions for beam shaping elements.

1. Introduction

   Radiation with ultra-high dose rate has recently attracted widespread attention and interest due to the so-called FLASH effect (Hornsey & Alper, 1966). Such an effect is a phenomenon by which high dose rate radiation rates (> 40 Gy/s) reduce normal tissue toxicity and maintain tumor control for the same dose as conventional (CONV) radiation (Favaudon, et al., 2014). The mechanism by which the normal tissues are spared has not been convincingly explained hitherto. The most commonly accepted explanation of this effect is oxygen depletion (Hornsey & Alper, 1966). It is theorized that ultra-fast dose rate radiation depletes oxygen in normal tissues necessary to "fix" the radiation-induced damage rendering them radio-resistant. However, a recent study (Jansen, et al., 2021) showed that FLASH radiation of different types (x-rays, protons and carbon ions) does not consume all of the oxygen. The experiments utilized a sealed 3D-printed water phantom exposed to 10 Gy dose and demonstrated that the amount of oxygen is not low enough to cause radio-resistance. The authors concluded that oxygen depletion may not be the only cause of improved normal tissue sparing effect of FLASH. Another hypothesis of the FLASH effect is related to the redox and free radicals. Normal tissue has more regular structure which can metabolize radicals more efficiently than cancerous cells. This effect can also be caused by radical-radical recombinations due the high density of ions produced at FLASH dose rates (Acharya, et al., 2011). More experiments and simulations are required to achieve a better understanding of FLASH effect.

Though the mechanism of the improved normal tissue sparing effect is uncertain, this phenomenon has been observed in different experiments performed by independent research teams. Field and Bewley (Field & Bewley, 1974) found that the radiation effectiveness in rat foot skin notably decreases when the dose rate increases to 83 Gy/s with 7 MeV electron beams. In 2014, Favaudon et al. (Favaudon, et al., 2014) found that FLASH was as efficient as CONV dose-rate irradiation in mice lung tumor treatment but no complications developed in normal tissues. Reduced incidence and severity of mouse skin ulceration were observed using FLASH radiation (180 Gy/s) compared to CONV x-ray radiation (0.0747 Gy/s) in a recent study (Soto, et al., 2020). Cognitive dysfunction is usually presented among brain tumor patients who received radiation treatment (Makale, McDonald, Hattangadi-Gluth, & Kesari, 2017). FLASH effect was also observed in brain radiation pre-clinical studies (Montay-Gruel, et al., 2017) (Montay-Gruel, et al., 2019) (Alaghband, et al., 2020). In a total abdominal irradiation, Levy et al. (Levy, et al., 2020) found that FLASH electron irradiation caused less gastrointestinal toxicity compared to CONV irradiation. The FLASH effect was also observed in higher mammals (mini pig and cat patients) (Vozenin, et al., 2019).

Almost all of the FLASH studies have focused for pre-clinical investigations, but some clinical trials were conducted in the past few years. The first clinic use of FLASH radiation was for a 75-year-old patient with cutaneous lymphoma (Bourhis, et al., 2019) . The reactions on normal tissue were mild and recovered sooner than that observed after conventional radiation. Meanwhile, rapid, long-lasting and complete tumor response was achieved with a short follow-up of 5 months.

Most of the FLASH irradiation is generated with electrons. Compared with electrons, protons can penetrate deeper for deep seated tumors. Recent studies proved the feasibility for FLASH proton beams. Varian and the Cincinnati Children's/UC Health Proton Therapy Center started human clinical trial for bone cancer treated with pencil beam scanning FLASH in November 2020 [AAPM talk]. And the output proved the FLASH effect.

A current challenge for proton FLASH is to how to generate beams with wide lateral profiles and spread-out Bragg peaks (SOBPs) to irradiate the target area large enough for clinical applications. The purpose of this study is to introduce automatic algorithms to optimize passive scattering proton beams at University of Texas MD Anderson Cancer Center Proton Therapy Center to obtain high dose rates utilizing a fast Monte Carlo (FDC) (Yepes, Randeniya, J, & Newhauser, 2009) (Yepes, Randeniya, Taddei, & Newhauser, 2009) (Yepes, et al., 2010).

2. Method

We use phase space files to model beams after the exit of the beam head in order to avoid simulating the beam-head and to significantly speed up the calculation of dose distributions. Then we optimize the FLASH proton beam by determining the collimator's radius, optimizing the flattening filter dimensions, compensator profile and ridge filter (RF) dimensions in sequence (Figure 1). A collimator is applied to reduce the penumbra of lateral dose profiles. And a flattening filter is to make the lateral dose profile flat and wide. We place a compensator inside of the collimator to fine adjust the range of the dose distribution. The collimator, flattening filter and compensator mainly shape the lateral dose profile. A RF is designed to spread out the Brag-peak in depth dose distribution.

The database of FDC uses water as a medium to save stopping powers, scattering angles and scaling parameters for other materials. However, water does not include all elements that a high-density material has, which may lead to dose calculation inaccuracy using FDC. It's possible that the FDC does not perform very well for the setup of this study because the flattening filter has a lead cone. After aperture optimization with FDC, we recalculate the dose distribution with Geant4 and verify that the inaccuracy of FDC is still within tolerance.

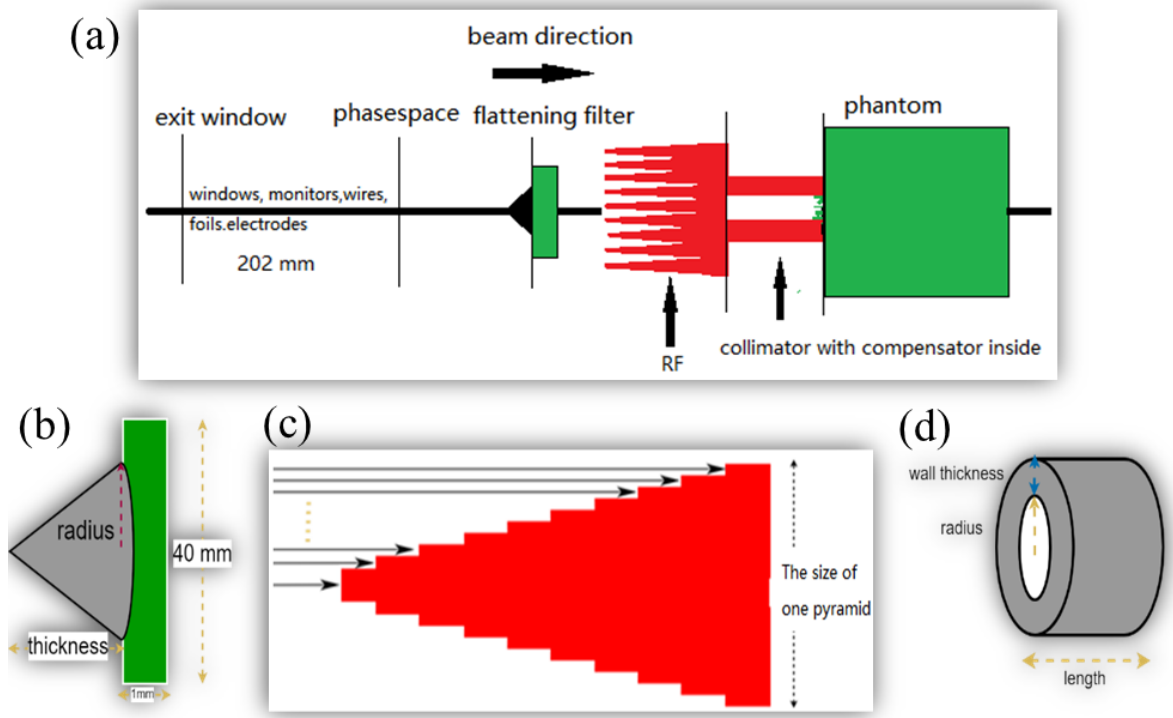

*Figure 1 Panel (a) displays the experiment setup. Panel (b), (c) and (d) are the flattening filter (a lead cone attached to a 1 mm Lucite plate), one pyramid of the RF and the collimator with optimized compensator inside (compensator can not be seen from this view), where dimension names are marked.*

(1) Phase space file generation

Before optimizing a FLASH beam with FDC, the first step is to generate a phase space file for each beam. The phase space file saves physical information (energies, spatial distribution and travelling directions) of one million particles passing through a plane (x-y plane) perpendicular to the beam direction (z-axis) located 202 mm downstream the beam nozzle exit window.

The phase space file is generated by tuning 6 parameters (Wang, et al., 2018) (Wang, et al., 2020), which are real energy, energy distribution width, spatial distribution widths in x and y axes and cosine of the angle between particle's travelling direction and x or y axis distribution widths. We assume that the energy, spatial and cosine of opening angle distributions are all Gaussians. The real energy and energy distribution width are the center and full width at half maximum (FWHM) of the Gaussian distribution. The spatial and cosine of opening angle distributions are centered on 0 and the FWHMs are variables that can be tuned to change the phase space file.

The six parameters are tuned to match FDC simulations with either experimental data or calculations fully simulated by MCNPX (Waters, et al., 2007). Figure 2 displays the FDC and MCNPX calculated integral depth dose (IDD) distribution and lateral dose profile along the x axis in water phantom at 5 different z locations for the energy 159.5 MeV as an example. Using the gamma index analysis technique, the passing rates for the two kinds of simulations are above 99.9% with 1%/1 mm index criterion for all these comparisons, which confirm the accuracy of the generated phase space file.

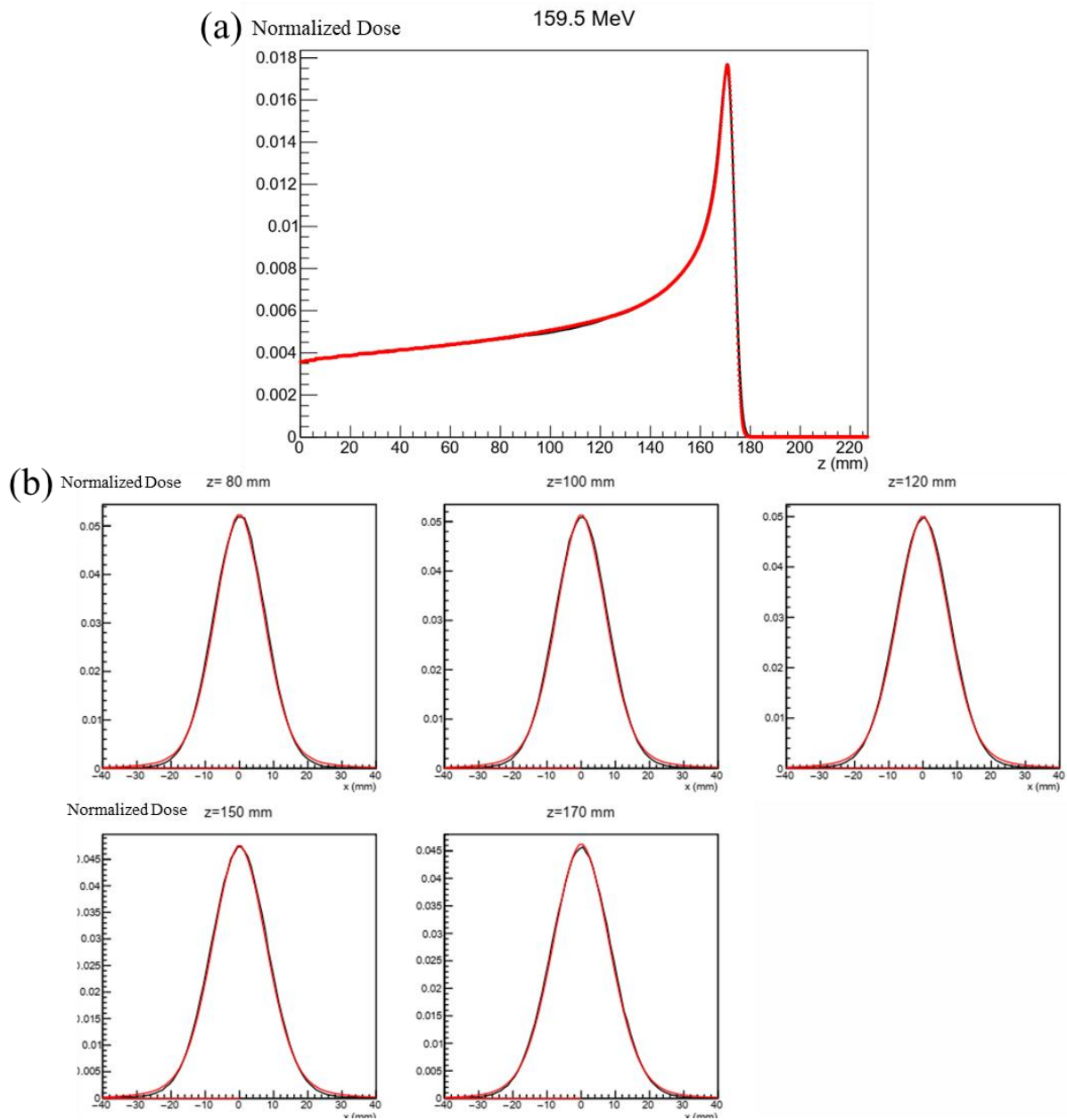

*Figure 2 Panel (a) is the IDD and Panel (b) shows the lateral dose profiles in x axis at 5 different depths for the energy 159.5 MeV. Black curves are for the FDC calculations and red ones are for the MCNPX simulations.*

(2) Collimator radius

We placed a brass collimator upstream of the water phantom to reduce penumbra of the lateral dose profile. The collimator should be long enough to remove the dose deposited by secondary particles produced by interaction of protons with the collimator on phantom. In this paper, the length selected for the energy 86.4 MeV and 159.5 MeV are 20 mm and 40 mm respectively. The collimator is used to reduce the penumbra but not to change the main shape of the lateral profile. Thus, we determine the radius of the collimator ($R_c$) by only inserting the flattening filter (Figure 1) in the beam path. By increasing the radius of the lead cone, the lateral dose profile at the Bragg-Peak changes from a Gaussian distribution to a shape with "horns" (beam intensity increases away from the central axis). When the horn magnitude reaches the maximum, the half width at 75% maximum of the dose profile is assigned as the collimator radius. We also record the position ($R_p$) of the two peaks at each side of the central axis, which will be used in the next step.

(3) Flattening filter dimensions

A lead cone attached to a 1 mm Lucite plate is used to flatten the peak of the lateral dose profile. The optimal dimensions of the cone change according to the beam and location. We optimized the radius of the cone by minimizing fluctuations of the lateral dose profile within the range (-$R_p$ ~ $R_p$). The standard deviation of the dose over the range -$R_p$ ~ $R_p$ is used to describe the level of fluctuations:

$$\bar{D} = \frac{1}{m}\sum_{j=1}^{m} D_j,$$

$$\sigma = \sqrt{\frac{1}{m}\sum_{j=1}^{m}(D_j - \bar{D})^2}$$, where j is coordinate index, $D_j$ is the dose in the $j^{th}$ bin

(4) Compensator profile optimization

We need a compensator to adjust dose ranges of different channels along z in the x-y plane. The compensator lateral thickness profile (Figure 3 Panel (a)) is determined by the material of the compensator and ranges of different channels. Because the proton beam and other beam shaping elements are symmetric around the z-axis, we only need to check ranges of channels along a radius in the x-y plane. The radius we select is along the positive x-axis. The material used for the compensator is Lucite, channel size along the radius is 0.5 mm and the range is defined to be 50% of the maximum dose along the specific channel. The compensator lateral thickness is calculated by:

$T_i = \frac{water\ density}{Lucite\ density} \times (z_i - z_0) + 1\ mm$, where $z_0$ is the range of the channel at x=0, and $z_i$ is the range of the $i^{th}$ channel.

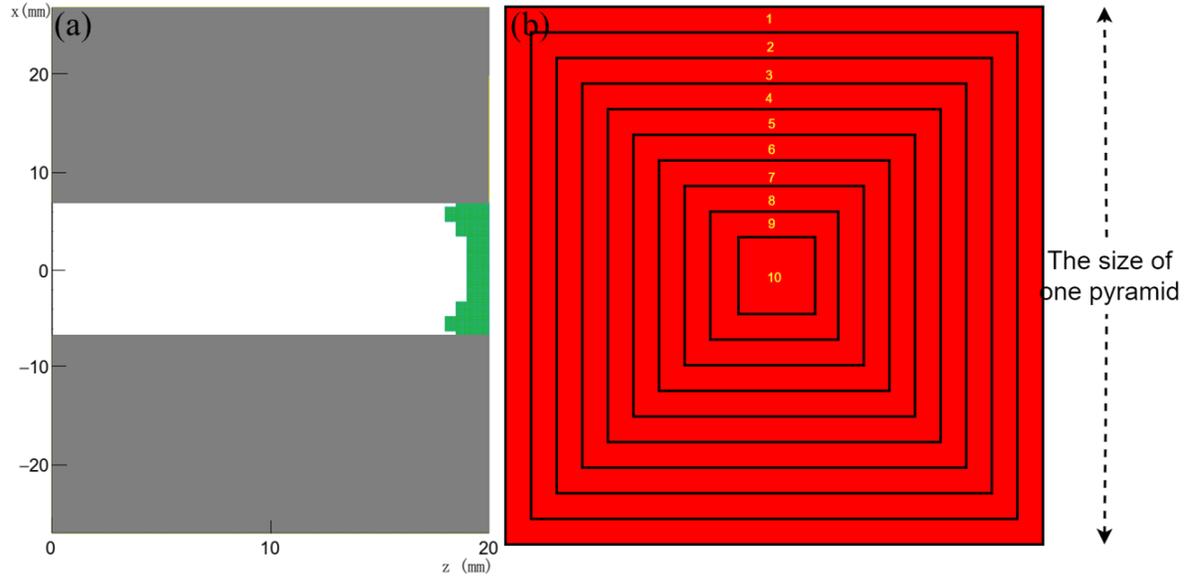

*Figure 3 Panel (a) is the longitudinal section of a collimator (grey) with optimized compensator (green) inside. Panel (b) is the top view of a pyramid. The area from edge to center is marked as 1, 2, 3 ....., 10.*

(5) RF dimension optimization

A RF is inserted to spread out the Bragg-peak of the depth dose distribution to cover the whole lengthways of the tumor. The design of the RF applies periodic pyramid filter with dimension of 2 mm x 2 mm x 10 mm for the energy 86.4 MeV and 4 mm x 4 mm x 40 mm for the energy 159.5 MeV. For each pyramid, there are 10 equally thick plates with decreasing widths stacked from bottom to top (Figure 1 Panel c).

The proton beam goes through different thicknesses of the pyramid material which results in their Bragg-peaks (BP) being distributed at different depths. The width of the SOBP is determined by the BP locations of the particles going through the thickest and thinnest pyramid materials. The flatness of the SOBP is determined by the number of particles going through different thicknesses. We adjust the weights of the 10 dose components to minimize the dose fluctuations within the SOBP.

The limited memory BSGS (L-BSGS) algorithm (Liu & Nocedal, 1989) is applied to optimize the dose component weights. The cost function is proportional to the standard deviation of dose values within the range of SOBP. The range is from peak inserting just one plate (the deepest point, $Z_d$) and 10 plates (the shallowest point, $Z_s$) in the RF.

The dose is expressed in terms of the summation of each individual dose component multiplied by its weight:

$D_j = \sum_{i=1}^{n} w_i d_{ij}$, where $i$ is the index of individual dose distribution with narrow Bragg-Peak, $j$ is the index of z position, n is 10 in this study, $d_{ij}$ is dose component from $i^{th}$ dose distribution with narrow peak at $j$ position and $w_i$ is the weight of the $i^{th}$ dose component.

The cost function is expressed as:

$$C = \frac{100\%}{\bar{D}} \sqrt{\frac{1}{k-1} \sum_{j=1}^{k} (D_j - \bar{D})^2}, \text{ where } \bar{D} = \frac{1}{m} \sum_{j=1}^{m} D_j.$$ Both $k$ and $m$ are numbers of position points. The $m$ is the total number of points within the range $Zs<z<Zd$. and $k$ is total number of points that dose is above $\bar{D}$ in the IDD.

And the derivative of the cost function is another important input to the L-BSGS optimization algorithm:

$$\frac{dC}{dw_i} = \frac{100\%}{\sqrt{k-1}} \frac{1}{\bar{D}^2} \left( \frac{\sum_{j=1}^{k} (D_j - \bar{D})(d_{ij} - \bar{d}_i)}{\sqrt{\sum_{j=1}^{k} (D_j - \bar{D})^2}} \bar{D} - \sqrt{\sum_{j=1}^{k} (D_j - \bar{D})^2} \, \bar{d}_i \right), \text{ where } \bar{d}_i = \frac{1}{m} \sum_{j=1}^{m} d_{ij}.$$

With the optimized weights for different components, we convert them to side lengths ($a_i$) of the 10 plates. The area of the $i^{th}$ ladder in one pyramid (**Error! Reference source not found.** Panel (b)) is proportional to the weight $w_i$:

$$\begin{cases} a_i^2 - a_{i+1}^2 = k * w_i, & i < 10 \\ a_i^2 = k * w_i, & i = 10 \end{cases}.$$

The parameter k can be obtained by summing all the 10 equations:

$$\begin{aligned}
k*w_{10} &= a_{10}^2 \\
k*w_9 &= a_9^2 - a_{10}^2 \\
k*w_8 &= a_8^2 - a_9^2 \\
&\vdots \\
k*w_2 &= a_2^2 - a_3^2 \\
+ \quad k*w_1 &= a_1^2 - a_2^2 \\
\hline
k*\Sigma w_i &= a_1^2
\end{aligned}$$

, $k = a_1^2 / \Sigma w_i$. $a_1$ is the size of one pyramid which is known and other $a_i$s can be obtained by plugging k into the 10 equations.

(6) FDC calculation validation
After all steps listed above are completed, we inserted those optimized beam shaping elements into Geant4 algorithm to have dose distribution recalculated. Gamma index analysis is applied to verify the accuracy of FDC by comparing with Geant4 simulations.

3. Results discussion

In this section, we present and discuss the optimized beams for the 86.4 MeV and 159.5 MeV energies when the water phantom is 550 mm downstream of the source (SSD = 550 mm). At the same time, we discussed the comparison between Geant4 and FDC simulations for both energies. We also moved the water phantom away from the source from 550 mm to 1000 mm, 1496 mm and 2000 mm respectively for the energy 159. 5 MeV and checked the re-optimized beam characters. The last part of this section is the dose rate estimation.

To describe beam characters, we discussed several quantities which are range, SOBP width, distal falloff, the 90% width, 80% width, 50% width and penumbra for the lateral dose profile. Range is the distance from surface to distal 90% of depth dose. SOBP width is the distance from proximal 90% to distal 90% of depth dose. Distal falloff is the distance from distal 80% to distal 20% of depth dose. The lateral 90% width, 80% width and 50% width are lateral width with dose above 90%, 80% and 50% respectively. The penumbra is defined as the distance from 80% to 20% of the maximum dose.

(1). Optimized beam characters and different simulations comparison

We present the two optimized beams with energy 86.4 MeV and 159.5 MeV in Figure 4 and 5 in which the distance between the phase space file and exit widow is 202 mm. The range for the optimized beam with energy 86.4 MeV is 51.8 mm. It has 8.5 mm SOBP width and the distal falloff is 2.5 mm. The lateral width is 11.0 mm, 12.0 and 14.5 mm with dose above 90%, 80% and 50% of maximum dose respectively. And the penumbra is 3.0 mm (Table 2).

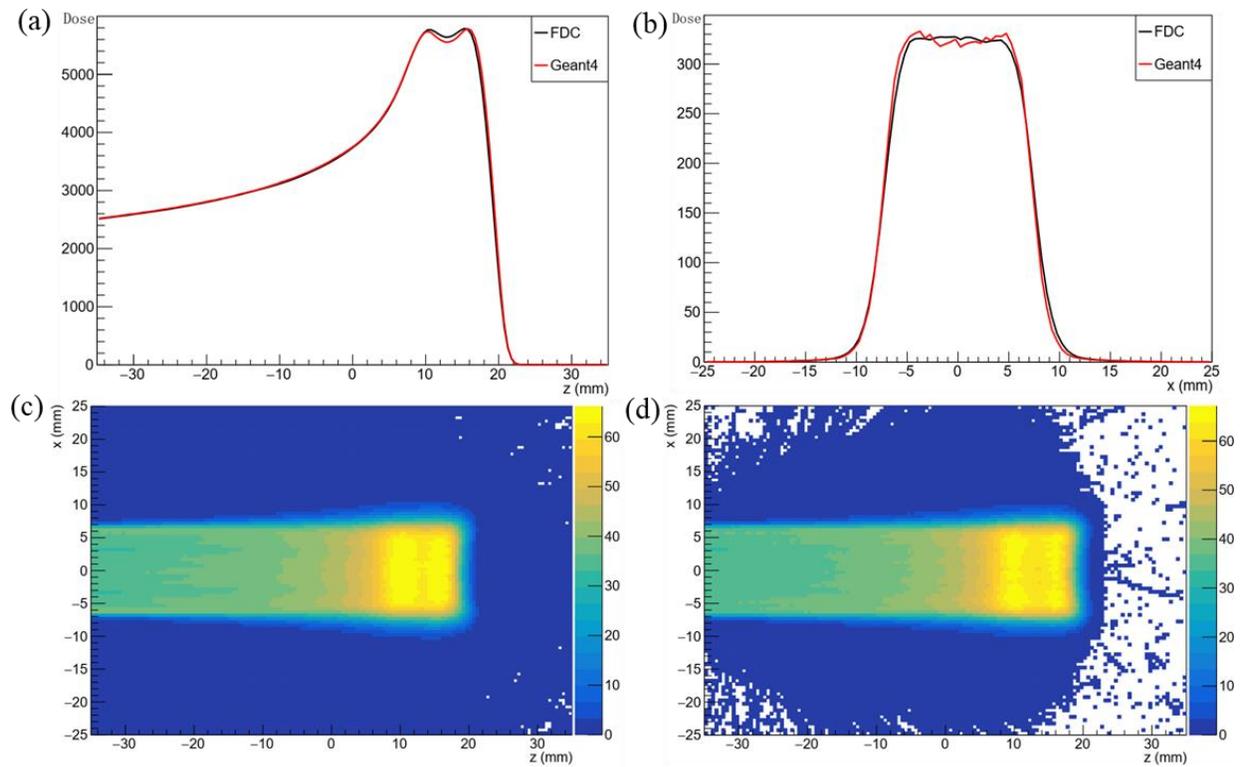

*Figure 4 Panel (a) is the integrated dose distribution for the energy 86.4 MeV. Panel (b) is the x profile for the energy 86.4 MeV. In both (a) and (b), black line is the calculation of Geant4 and red line is from FDC. Panel (c) and (d) are x-z cross section with y integrated from -2 mm to 2 mm for FDC and Geant4 respectively.*

We also listed quantities for the optimized beam with energy 159.5 MeV in Table 2. The range, SOBP width and distal falloff of the IDD are 161.8 mm, 39.0 and 4.5 mm respectively. The lateral dose profile is with 5.75 mm penumbra, and the 90%, 80% and 50% lateral widths are 12.5 mm, 15.0 mm and 20.5 mm respectively.

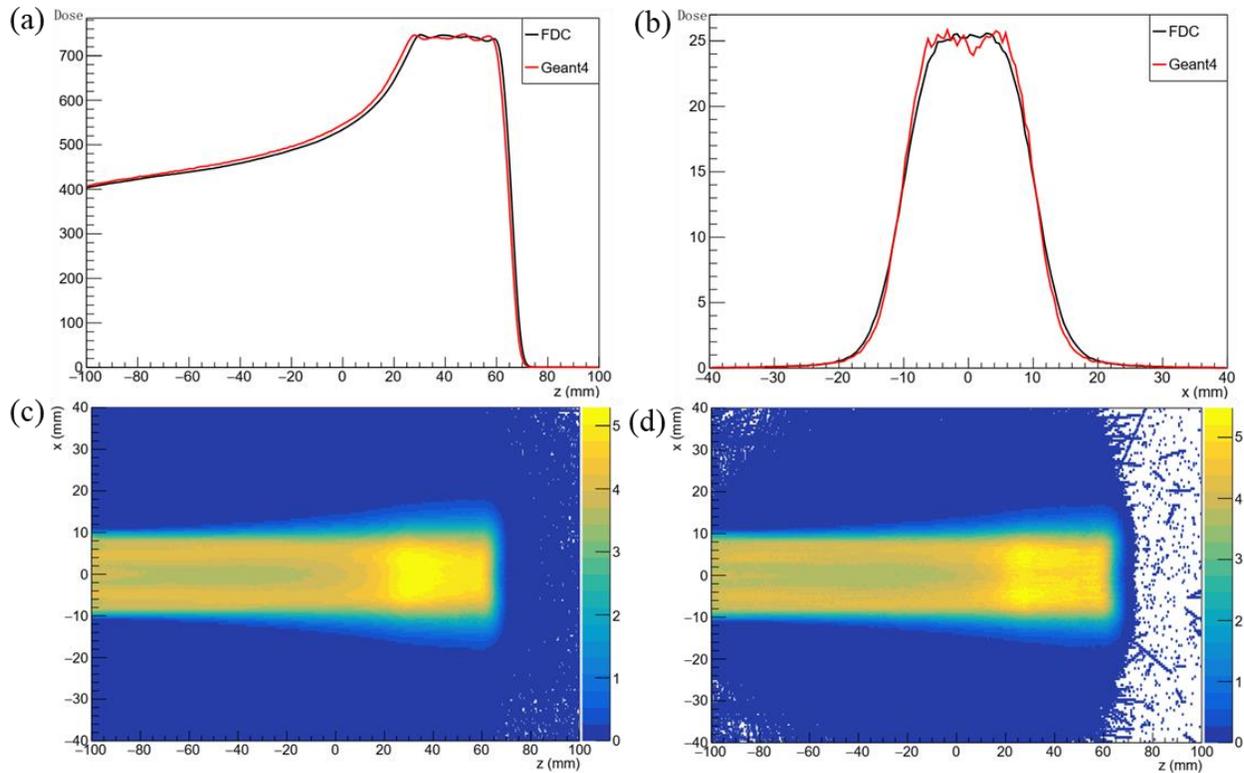

*Figure 5 Plot (a) is the integrated dose distribution for the energy 159.5 MeV. Plot (b) is the x profile for the energy 159.5 MeV. In both (a) and (b), black line is the calculation of Geant4 and red line is from FDC. Panel (c) and (d) are x-z cross section with y integrated from -2 mm to 2 mm for FDC and Geant4 respectively.*

| Energy (MeV) | Range (mm) | SOBP (mm) | Distal Falloff (mm) | Lateral 90% width (mm) | Lateral 80% width (mm) | Lateral 50% width (mm) | Penumbra (mm) |
|---|---|---|---|---|---|---|---|
| 86.4 | 51.8 | 8.5 | 2.5 | 11.0 | 12.0 | 14.5 | 3.0 |
| 159.5 | 161.8 | 39.0 | 4.5 | 12.5 | 15.0 | 20.5 | 5.75 |

*Table 1 Optimized beam characters for the energy 86.4 MeV and 159.5 MeV respectively when SSD is 550 mm.*

The FDC calculations have a passing rate 99.9% for the energy of 86.4 MeV and 96.2% for 159.5 MeV, using 3%/3 mm as a gamma-index criterion, compared to Geant4 simulations. The lateral profile comparison is worse for the energy of 159.5 MeV because the lead size used for this energy is larger. The performance of FDC for high density material is not as accurate as that for low density material.

(2). Optimized beam with different SSDs

We moved the water phantom further away from the source to obtain larger field size. Figure 6 displays dose distributions of optimized beams for the energy 159.5 MeV with different SSDs. The Panel (a) and (b) are IDD and lateral dose profile in x axis when SSD is 550 mm, 1000 mm, 1496 mm and 2000 mm. The rest four plots in the same figure are 2D dose profile in the x-z plane with y integrated from -2 mm to 2 mm when SSD was set to be 550 mm (a), 1000 mm (b), 1496 mm (c) and 2000 mm (d) respectively.

In Table 2, we listed quantities of optimized beams with different SSDs for the energy 159.5 MeV. The optimized beam had SOBP width 39.0 mm for 550 SSD and 48.0 mm when SSD is 1000 mm, 1496 mm and 2000 mm. The SOBP does not change much for different SSDs because the SOBP is determined by the total thickness of RF. The range is 161.8 mm, 158.3 mm, 157.3 mm and 155.8 mm when SSD is 550.0 mm, 1000.0 mm, 1496.0 mm and 2000.0 mm respectively. The range decreases slightly because particles diverge before they hit the water phantom. The distal falloff is caused by fluctuations in the energy loss of individual protons (Newhauser & Zhang, 2016) which is about the same for the same beam with different SSDs. So the distal falloffs for different SSD are close in Table 2.

In the lateral dose profile plot, the optimized beam had 17.0 mm, 19.0 mm and 25.0 mm lateral width with doses above 90%, 80% and 50% of the maximum respectively for 1000 mm SSD. The lateral widths with dose above 90%, 80% and 50% of the maximum are 19.0 mm 23.0 mm and 29.0 mm respectively for the 1496 mm SSD. And for the 2000 mm SSD, the values are 23.0 mm, 26.0 mm and 32.0 mm. The lateral widths increase as SSD increases which makes sense because particles diverge along the beam direction. The lateral dose profile penumbra is 6.5 mm for 1000 mm, 1496 mm and 2000 mm SSDs and the value is 6.0 mm when SSD is 550 mm. The penumbras are close for different SSDs because the collimator radius was determined by the width with doses above 75% in lateral dose profile only with flatter filter placed in beam path.

| SSD (mm) | Range (mm) | SOBP (mm) | Distal falloff (mm) | Lateral 90% width (mm) | Lateral 80% width (mm) | Lateral 50% width (mm) | Penumbra (mm) |
|---|---|---|---|---|---|---|---|
| 550.0 | 161.8 | 39.0 | 4.5 | 13.0 | 15.0 | 21.0 | 6.0 |
| 1000.0 | 158.3 | 38.0 | 5.5 | 17.0 | 19.0 | 25.0 | 6.5 |
| 1496.0 | 157.3 | 38.5 | 5.0 | 19.0 | 23.0 | 29.0 | 6.5 |
| 2000.0 | 155.8 | 38.0 | 5.5 | 23.0 | 26.0 | 32.0 | 6.5 |

*Table 2 Optimized beam characters for different SSDs for the energy 159.5 MeV.*

The lateral width increases with increasing SSD. The IDD decreases with increasing SSD except the 550 mm SSD. The IDD decreases because more dose is absorbed by beam shaping elements. The elements contributing the most to that dose absorption are the lead cone and collimator. For a larger SSD, a larger radius is required for the lead cone to obtain a flat lateral dose profile and more particles are scattered by the lead cone. The 550 mm SSD case has a lead cone with a shorter radius, however, the narrow collimator causes more scatters which lower down the IDD curve.

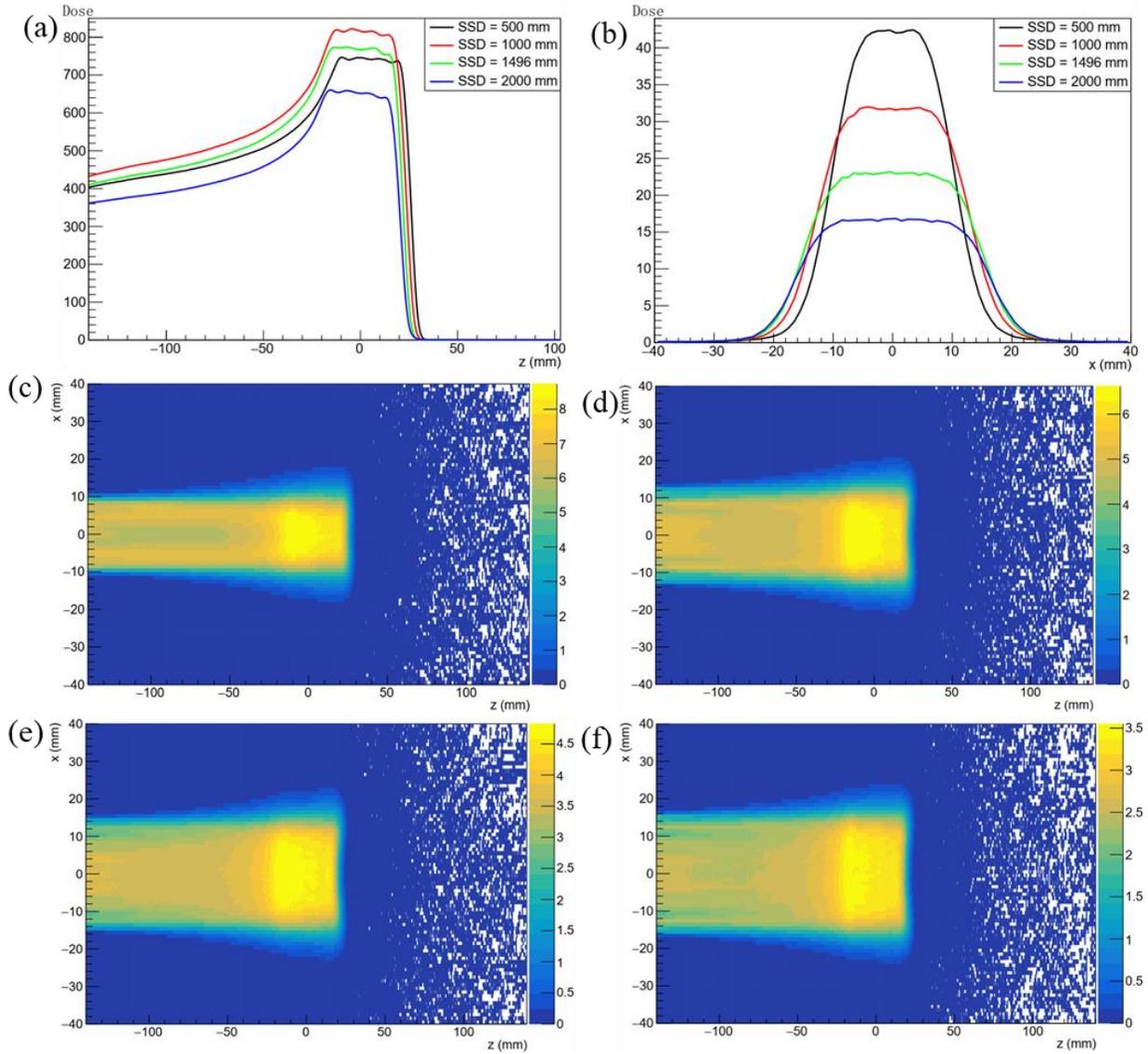

*Figure 6. Dose distributions of optimized beams for the energy 159.5 MeV. Panel (a) and (b) are the IDD and the lateral dose profile in x axis for different SSDs. The rest panels in the same figure are 2D dose profile in the xz plane with y integrated from -2 mm to 2 mm when SSD was set to be 550 mm (a) to 1000 mm (b), 1496 mm (c) and 2000 mm (d) respectively.*

(3). Dose rate

Dose rate is the most important quantity for the flash radiation investigation. Because the optimization discussed above is for one single beam, the dose rate is not of concern during the optimization process. After the optimization, we need to estimate the dose rate to ensure it is greater than the threshold for the flash effect. The method applied for the dose rate estimation is to scale the 3D dose distribution with beam shaping elements to the dose distribution without any modifiers in the beam path (original beam). For example, if the entrance dose rate is 100 Gy/s for the energy 159.5 MeV, the entrance dose rate for the optimized beam is 150 Gy×scale. The entrance dose is 6.43, 5.17, 4.20, 2.93 and 2.02 for the original beam, the optimized beam when SSD is 500 mm, 1000 mm, 1496 mm and 2000 mm respectively (Figure 7). The dose scales are 0.80, 0.65, 0.46 and 0.31 respectively.

And the entrance dose rate is 120 Gy/s, 97.5 Gy/s, 69 Gy/s, 46.5 Gy/s for the optimized beam with SSD 500 mm, 1000 mm, 1496 mm and 2000 mm.

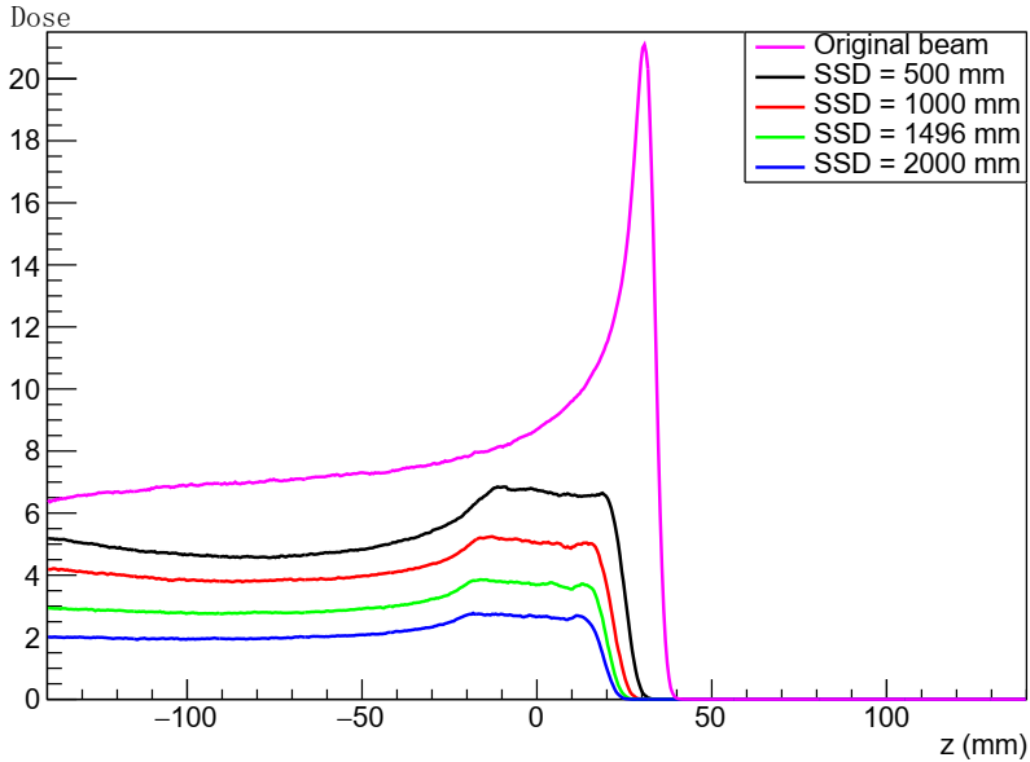

*Figure 7 The depth dose profile in the center for the energy 159.5 MeV. The entrance dose is 6.43, 5.17, 4.20, 2.93 and 2.02 for the original beam, the optimized beam when SSD is 500 mm, 1000 mm, 1496 mm and 2000 mm.*

4. Conclusion

A set of algorithms was introduced to automatically optimize dimensions of flattening filter, RF, collimator and compensator for FLASH proton beam using FDC, a track-repeating fast Monte Carlo method. The optimization process includes six steps in total, which are (1) phase space file generation for each beam, (2) collimator radius determination, (3) radius selection for lead cone in flattening filter, (4) compensator profile determination, (5) RF dimension optimization and (6) FDC result verification through comparing with Geant4. We tested two energies, 86.4 MeV and 159.5 MeV with our optimizing algorithms. And for the energy 159.5 MeV, four SSDs (550 mm, 1000 mm, 1496 mm and 2000 mm) were explored. Through checking ranges, SOBP widths, distal falloffs, lateral dose profile widths and penumbras for different optimized beams, we conclude that high beam quality can be obtained using our optimization algorithms. After optimization, we determined the dose rate for an optimized beam by scaling it to the dose distribution of the original beam.

**ACKNOWLEDGMENTS**
We are grateful to Mr. Yupeng Li for critical discussion on beam shaping method.